\documentstyle[preprint,aps]{revtex}

\begin{document}
\draft
\preprint{HEP/123-qed}
\title{A sandpile model with tokamak-like enhanced confinement
phenomenology}
\author{S. C. Chapman$^1$\footnote{sandrac@astro.warwick.ac.uk},
}
\address{
$^1$Physics Dept. Univ. of Warwick,
Coventry CV4 7AL, UK
}
\author{
R. O. Dendy$^2$}
\address{ 
$^2$EURATOM/UKAEA Fusion Association, Culham Science Centre, Abingdon,
Oxfordshire OX14 3DB, United Kingdom}
\author{B. Hnat$^1$}
\date{\today}
\maketitle
\begin{abstract}

Confinement phenomenology characteristic of magnetically confined plasmas
emerges naturally from a simple sandpile algorithm when the parameter
controlling redistribution scalelength is varied. Close analogues are found
for enhanced confinement, edge pedestals, and edge localised modes (ELMs),
and for the qualitative correlations between them. These results suggest
that tokamak observations of avalanching transport are deeply linked to
the existence of enhanced confinement and ELMs.

\end{abstract}
\pacs{
52.55.Dy, 52.55.Fa, 45.70.Ht, 52.35.Ra}

The introduction of the sandpile paradigm\cite{btw87}-\cite{Jensen98}
into magnetized plasma physics
(fusion\cite{Newman:Carreras96}-\cite{Politzer2000},
magnetospheric\cite{Chang92}-\cite{Lui:Aprilfirst},
and accretion disk\cite{Mineshige:Takeuchi}-\cite{Dendy:Tagger}: for recent
reviews see Ref.\cite{chap:dend:row99,ssr}) has opened new conceptual avenues.
It provides a framework within which observations of rapid nondiffusive nonlocal
transport phenomena can be studied; recent examples include analyses of
auroral energy deposition derived from global imaging\cite{Lui:Aprilfirst},
and of electron temperature fluctuations in the DIII-D
tokamak\cite{Politzer2000},
both of which involve avalanching. Insofar as such phenomena resemble
those in experimental sandpiles or mathematically idealized models thereof, they
suggest that the confinement physics of macroscopic systems (plasma and
other) may reflect unifying underlying principles.

In this paper we present results suggesting that this unity may extend to
some of the most distinctive features of toroidal magnetic plasma confinement:
enhanced confinement regimes (``H-modes"), edge localised modes (``ELMs"), steep
edge gradients (``edge pedestal"), and their observed phenomenological and
statistical correlations -- for recent quantitative studies, see for example
Refs.\cite{Fishpool98,Zhang98,hugill2000} and references therein. An important
question is whether the L to H transition necessarily reflects a catastrophic
bifurcation of confinement properties, or can be associated with a monotonic
change in the character of the turbulence\cite{hugill2000}.Transitional behaviour resembling that observed in fusion
plasmas has been found in other sandpile
algorithms: for example in Ref \cite{Newman:Carreras96}
a local sheared flow region is imposed
which reduces the characteristic avalanche length thereby affecting confinement;
and in Ref \cite{carerras:stiffness}, changes in the redistribution rule lead to changes in
profile stiffness. We show that key
elements of the observed phenomenology emerge naturally
from a simple one-dimensional sandpile model, that of Chapman
\cite{Chap:Row:2000}, which incorporates other
established models\cite{btw87,dendy98} as limiting cases.
This centrally fueled (at cell $n = 1$)  model's distinctive
algorithmic feature relates to the local redistribution of sand at a cell
(say at $n = k$) where the critical gradient $z_c$ is exceeded: the sandpile is
instantaneously flattened behind the unstable cell over a 
length $L_f$, embracing the cells $n = k-(L_f-1),k-(L_f-2),...,k$;
and this sand is conservatively relocated to the cell at $n = k + 1$.
$L_f$  defines the length scale over which the most rapid 
redistribution occurs; in a plasma context this could be considered a proxy for turbulent correlation length or
eddy size.
In Ref.\cite{Chap:Row:2000} the sandpile is explored for all regimes $1
< L_f < N$
for both constant and fluctuating critical gradient $z_c$. Here
we consider the dynamics of the more realistic case with random
fluctuations in $z_c$;
the system is robust in that once fluctuations are introduced in the critical
gradient, the behavior is essentially insensitive to both their level
and spectral properties\cite{Chap:Row:2000}, see also Ref.\cite{chap:dend:row99}.
The limit $L_f = 1$ is the  fixed point corresponding to
the centrally fueled algorithm of Ref.\cite{btw87} in one dimension. In the
limit $L_f = N$ (where $N$ is the number of cells in the sandpile) the
sandpile is
flattened everywhere behind an unstable cell as in
Refs.\cite{dendy98,chap:dend:row99}.
A real space renormalization group approach\cite{tam}
shows that the robust scale free dynamics for the limiting case $L_f=N$
corresponds to a nontrivial (repulsive) fixed point (see e.g. Ref.\cite{Jensen98}).
The essential result of  Ref.\cite{Chap:Row:2000} is that different regimes
of avalanche statistics emerge, resembling a transition from regular to
intermittent
dynamics reminiscent of deterministic chaos. The control parameter
is the normalized redistribution scalelength  $L_f/N$ which specifies whether
the system is close to the nontrivial $L_f=N$ fixed point.

Height profiles for the sandpile with 512 cells, time averaged over
many thousands of avalanches, are shown in Fig.1
for three different values of the fluidization length $L_f$
in the range $50 < L_f < 250$. The sandpile profile shape, stored
gravitational potential energy, and edge structure (smooth decline or
pedestal) correlate
with each other and with $ L_f$. As $L_f$ is reduced, the edge pedestal
steepens and the
time averaged stored energy rises; multiple ``barriers" (regions of steep
gradient)
are visible in trace (a) and to some extent trace (b) of Fig.1.
Time evolution of the sandpile for $L_f = 50, 150$, and 250
respectively is quantified
in Figs.2-4. The top traces show total stored energy; the middle traces show the
position of the edge of the sandpile (the last occupied cell); and the
bottom traces show the
magnitude and occurrence times of mass loss events (hereafter MLEs) in
which sand is lost
from the system by being transferred beyond the 512th cell. Time is normalized
to the mean inter-avalanche time $\Delta \tau$ (proportional to the fueling
rate). The sandpile is fueled only at the first cell, so that the great
majority of
avalanches terminate before reaching the 512th cell (these are classified
as internal).
While internal avalanches result in energy dissipation (recorded in the
upper traces of
Figs.2-4), and may alter the position of the edge of the sandpile, they do
not result
in an MLE; there are corresponding periods of quiescence in the middle and
lower traces of
Figs.2-4. Conversely the MLEs are associated with sudden inward movement of
the sandpile edge,
and in this important sense appear to be edge localised. However, MLEs and
the associated
inward edge movement are in fact the result of systemwide avalanches
triggered at the
sandpile center (cell $n=1$). The character of the MLEs changes with $L_f$.
In Fig.2, where
the mean and peak stored energy are greatest, the MLEs are similar to each
other and
occur with some regularity. The regularity of MLE occurrence in Fig.3 is
less marked,
the magnitude of the largest MLEs is greater than in Fig.2, and there is
greater spread in MLE size. This trend continues in Fig.4, which also has
the lowest stored energy. These effects correlate with the underlying
dynamics of the
sandpile. Figure 5 plots the relation between average stored energy and
$L_f$ for 
the $N=512$ system and  much larger $N=4096$ and $8192$ systems
(normalized to the
system size $N$).
The curves coincide, demonstrating invariance with respect to
system size, with an inverse
power law with slope
close to $-2$ for $ L_f/N < 1/4$, and a break at $L_f/N \sim 1/4$. These
two regimes yield
the quasi-regular and quasi-intermittent dynamics in Figs.2-4 (see also the
plot of avalanche
length distribution against $L_f$ in Fig.8 of Ref.\cite{Chap:Row:2000}).
The parameter $L_f/N$ is a measure of proximity of this high dimensional
system to the
$L_f=N$ nontrivial fixed point. This determines both the apparent
complexity of the
timeseries in Figs.2-4 and the underlying statistical simplicity described
below,
which is also invariant with respect to system size.

There is systematic correlation between time averaged stored energy $<E>$
and MLE frequency $f_{MLE}$, as shown in Fig.6. To obtain these curves,
which are again normalized to system size, we have derived MLE frequencies
using a standard algorithm previously used\cite{Zhang98} to assign
frequencies to ELMs observed in tokamak plasmas in the Joint European Torus
(JET).
Since the sandpile often generates bursts of mass loss with structure
down to
the smallest timescales, which might not be resolvable
under experimental conditions, we have followed Ref.\cite{Zhang98} in
applying a (relatively narrow) measurement window of width $450 \Delta \tau$
to obtain $f_{MLE}$. The correlation between $<E>$ and $f_{MLE}$ is a
noteworthy
emergent property, furthermore Fig.6's characteristic curve is very similar
to that
of Fig.6 of Ref.\cite{Fishpool98}, which relates measured energy confinement to
ELM frequency in JET. Energy confinement time $\tau_c$ can be defined for
the sandpile
by dividing the time averaged stored energy $<E>$  by the time averaged
energy dissipation rate $<\Delta E>$ (where $\Delta E$ is the energy
dissipated in a
single avalanche). The embedded plot of Fig. 6 shows $\tau_c$ against MLE
frequency $f_{MLE}$.

Finally, we explore the situation where there is a secular change
in the redistribution algorithm: in Fig.7, $L_f$ decreases slowly,
continuously, and linearly with time, from one constant value to another over
a period encompassing many tens of thousands of avalanches. There is a
corresponding time evolution of the energy confinement properties of the
sandpile
and of the character of the MLEs. Figure 7(top) shows total stored energy
as a function of time as $L_f$ changes from $250$ at $t=4\times 10^4$ to $50$
at $t=1.15\times 10^5$, while $\sim 10^5$ avalanches occur: over a period
of time
corresponding to a few tens of MLEs, the system smoothly evolves from low
to high confinement.
This is accompanied by a gradual change in character of the time variation
in the sandpile edge
(position of last occupied cell, Fig.7(middle)) and of the MLEs
(Fig.7(lower)), from large amplitude to small and from irregular to regular.
Figure 7 can perhaps be regarded as this  sandpile's analogue of, for example,
Fig.2 of Ref.\cite{Zhang98} or Fig.2 of\cite{hugill2000}. The essential
point here is that
the sandpile apparently freely explores phase space with changing control
parameter $L_f/N$.
Characteristic properties of the dynamics (whether quasi-regular or
quasi-intermittent) and
correspondingly, confinement properties (such as stored energy and MLE
characteristics)
smoothly follow changes in this parameter rather than exhibiting a sudden
phase transition
or catastrophe.

By varying a single control parameter in the sandpile algorithm, we
have shown correlations
between: stored energy, confinement times, sandpile profile, sandpile edge
structure, and the
amplitude, frequency, and dynamical character of mass loss events. We have
also seen how slow
secular change in the control parameter produces a smooth evolution in
confinement properties. If a single control parameter analogous to $L_f/N$
exists for tokamaks,
it can in principle be found
from experimental data by examining scaling with respect to system size 
as above. 

The existence of such extensive tokamak-like phenomenology, emergent from a very
simple system, is a novel discovery. Insofar as the phenomenological
resemblance is close,
there is more to be learnt. A minimalist interpretation starts from the
premise that the
sandpile algorithm of Ref. \cite{Chap:Row:2000} provides a simple one-parameter model for studying
generic nonlocal
transport, conditioned by a critical gradient, in a macroscopic confinement
system.
Changing the value of the single control parameter $L_f$ then corresponds
to altering
the range in configuration space over which the transport process operates.
It then
follows from the results in the present paper that this may be the minimum requirement to
generate those aspects of tokamak-like confinement phenomenology described.
This is a significant
conclusion, but one can consider a more far-reaching one. A possible
maximalist interpretation
attaches greater weight to recent
observations\cite{Carreras:vanM:98,rhodes99,Politzer2000}
of avalanching transport in tokamaks and in largescale numerical
simulations\cite{Garbet:Waltz,Sarazin:Ghendrih} thereof, and therefore
regards the avalanching
transport that is built into sandpile algorithms as an additional point of
contact
with magnetically confined plasmas. One would then infer from the present
results that
tokamak observations of avalanching transport are deeply linked to the
existence of enhanced
confinement and ELMs.

\acknowledgements
We are grateful to Jack Connor, George Rowlands, David Ward and Nick Watkins
for comments and suggestions. SCC was supported by a PPARC lecturer
fellowship, ROD by
Euratom and the UK DTI, and BH by HEFCE.

Captions

FIG.1.  Time averaged height profiles of the 512 cell sandpile
for $L_f =$ (a)50, (b)150 and (c)250. Inset: edge structure.

FIG.2.  Time evolution of the 512 cell sandpile with $L_f = 50$:
(top) stored energy, (middle) position of last occupied cell,
(lower) magnitude and occurence of mass loss events.

FIG.3.  As Fig.2, for $L_f = 150$.

FIG.4.  As Fig.2, for $L_f = 250$.

FIG.5.  Average stored energy versus $L_f/N$ for sandpiles of
$N=512, 4096, 8192$. Energy is normalized to the $L_f=1$ case (effectively to
$N^2$).

FIG.6.  Average stored energy versus MLE frequency, and (inset) $\tau_c$
versus MLE frequency
for sandpiles of
$N=512, 4096, 8192$. Energy and MLE frequency are normalized as in Fig 6.

FIG.7. Time evolution of (top) stored energy, (middle) sandpile edge
position and
(lower) MLEs, as $L_f$ changes slowly and linearly from 250 to 50.


\begin{references}

\bibitem{btw87}
P. Bak, C. Tang, and K. Wiesenfeld,
Phys. Rev. Lett. {\bf 50}, 381 (1987).

\bibitem{Bak:Tang}
P. Bak, C. Tang, and K. Wiesenfeld,
Phys. Rev. A {\bf 38}, 364 (1988).

\bibitem{Jensen98}
H.J. Jensen,
{\it Self-Organised Criticality:  Emergent Complex Behaviour in Physical
and Biological Systems}, Cambridge University Press, 1998.

\bibitem{Newman:Carreras96}
D.E. Newman, B.A. Carreras, P.H. Diamond, and T.S. Hahm,
Phys. Plasmas {\bf 3}, 1858 (1996).

\bibitem{new2} 
B. A. Carreras, D. Newman, V. E. Lynch and P. H. Diamond, Phys. Plasmas, {\bf 3}, 2903 (1996)

\bibitem{carerras:stiffness} 
B. A. Carreras, V. E. Lynch, P. H. Diamond and M. Medvedev, Phys. Plasmas,
{\bf 5}, 1206 (1998).

\bibitem{Dendy:Helander97}
R.O. Dendy and P. Helander,
Plasma Phys. Control. Fusion {\bf 39}, 1947 (1997).

\bibitem{Carreras:vanM:98}
B.A. Carreras {\it et al.},
Phys. Rev. Lett. {\bf 80}, 4438 (1998).

\bibitem{Garbet:Waltz}
X. Garbet and R. Waltz,
Phys. Plasmas {\bf 5}, 2836 (1998).

\bibitem{Sarazin:Ghendrih}
Y. Sarazin and P. Ghendrih,
Phys. Plasmas {\bf 5}, 4214 (1998).

\bibitem{rhodes99}
T.L. Rhodes {\it et al.},
Phys. Lett. A {\bf 253}, 181 (1999).

\bibitem{Politzer2000}
P.A. Politzer,
Phys. Rev. Lett. {\bf 84}, 1192 (2000).

\bibitem{Chang92}
T.S. Chang,
IEEE Trans. Plasma Sci. {\bf 20}, 691 (1992).

\bibitem{Chapman98}
S.C. Chapman, N.W. Watkins, R.O. Dendy, P. Helander, and G. Rowlands,
Geophys. Res. Lett. {\bf 25}, 2397 (1998).

\bibitem{Lui:Aprilfirst}
A.T.Y. Lui {\it et al.},
Geophys. Res. Lett. {\bf 27}, 2397 (2000).

\bibitem{Mineshige:Takeuchi}
S. Mineshige, M. Takeuchi, and H. Nishimori,
Astrophys. J. {\bf 435}, L125 (1994).

\bibitem{Leighly:O'Brien}
K.M. Leighly and P.T. O'Brien,
Astrophys. J. {\bf 481}, L15 (1997).

\bibitem{Dendy:Tagger}
R.O. Dendy, P. Helander, and M. Tagger,
Astron. Astrophys. {\bf 337}, 962 (1998).

\bibitem{chap:dend:row99}
S.C. Chapman, R.O. Dendy, and G. Rowlands,
Phys. Plasmas {\bf 6}, 4169 (1999).

\bibitem{ssr}
S.C. Chapman and N.W. Watkins,
Space Sci. Rev. accepted (2000).

\bibitem{Fishpool98}
G.M. Fishpool,
Nucl. Fusion {\bf 38}, 1373 (1998).

\bibitem{Zhang98}
W. Zhang, B.J.D. Tubbing, and D. Ward,
Plasma Phys. Control. Fusion {\bf 40}, 335 (1998).

\bibitem{hugill2000}
J. Hugill,
Plasma Phys. Control. Fusion {\bf 42}, R75 (2000).

\bibitem{Chap:Row:2000}
S.C. Chapman,
Phys. Rev. E {\bf 62}, 1905 (2000).

\bibitem{dendy98}
R.O. Dendy and P. Helander,
Phys. Rev. E {\bf 57}, 3641 (1998).

\bibitem{tam}
S.W.Y. Tam, T.S. Chang, S.C. Chapman, and N.W. Watkins,
Geophys. Res. Lett. {\bf 27}, 1367 (2000).

\end{references}
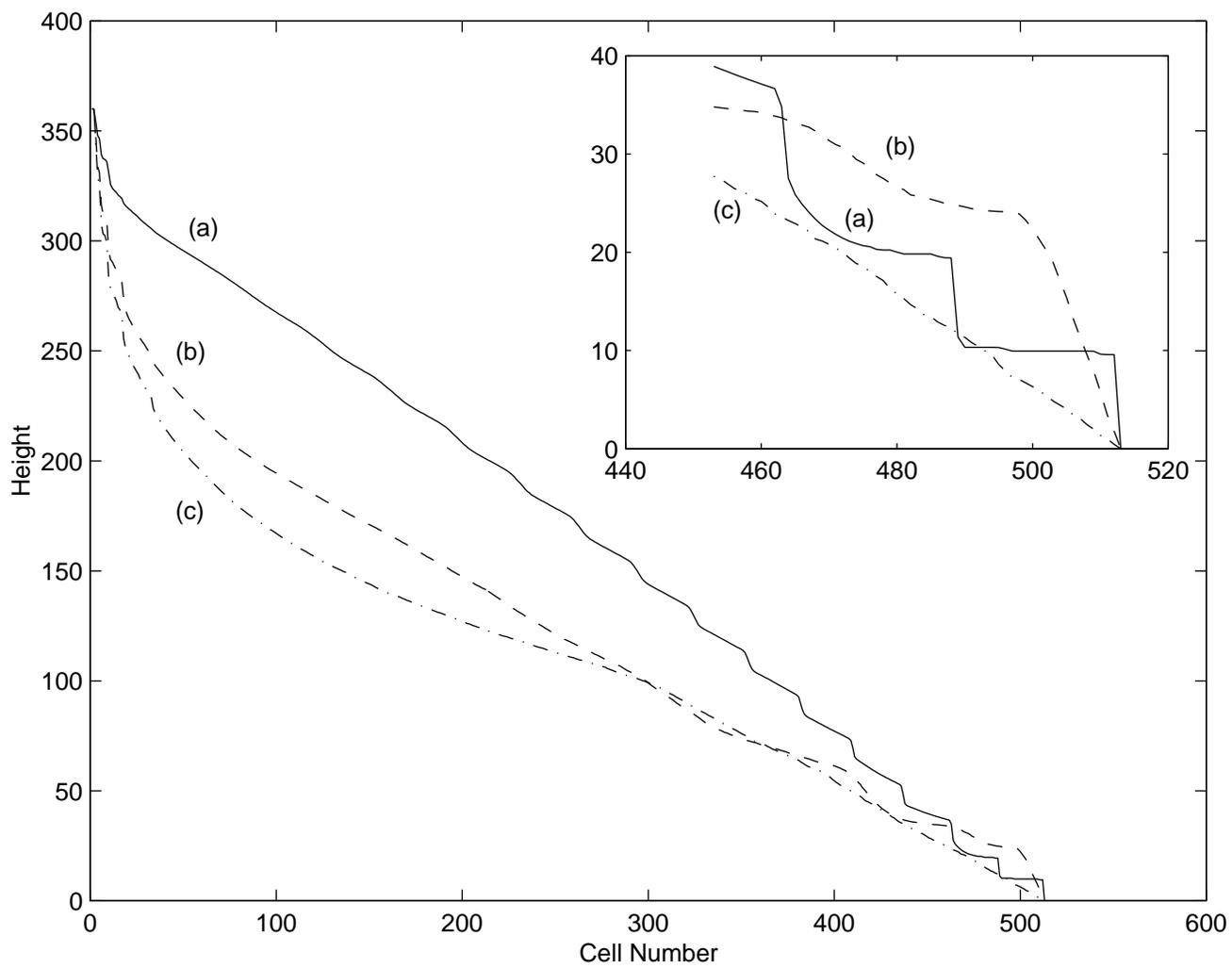
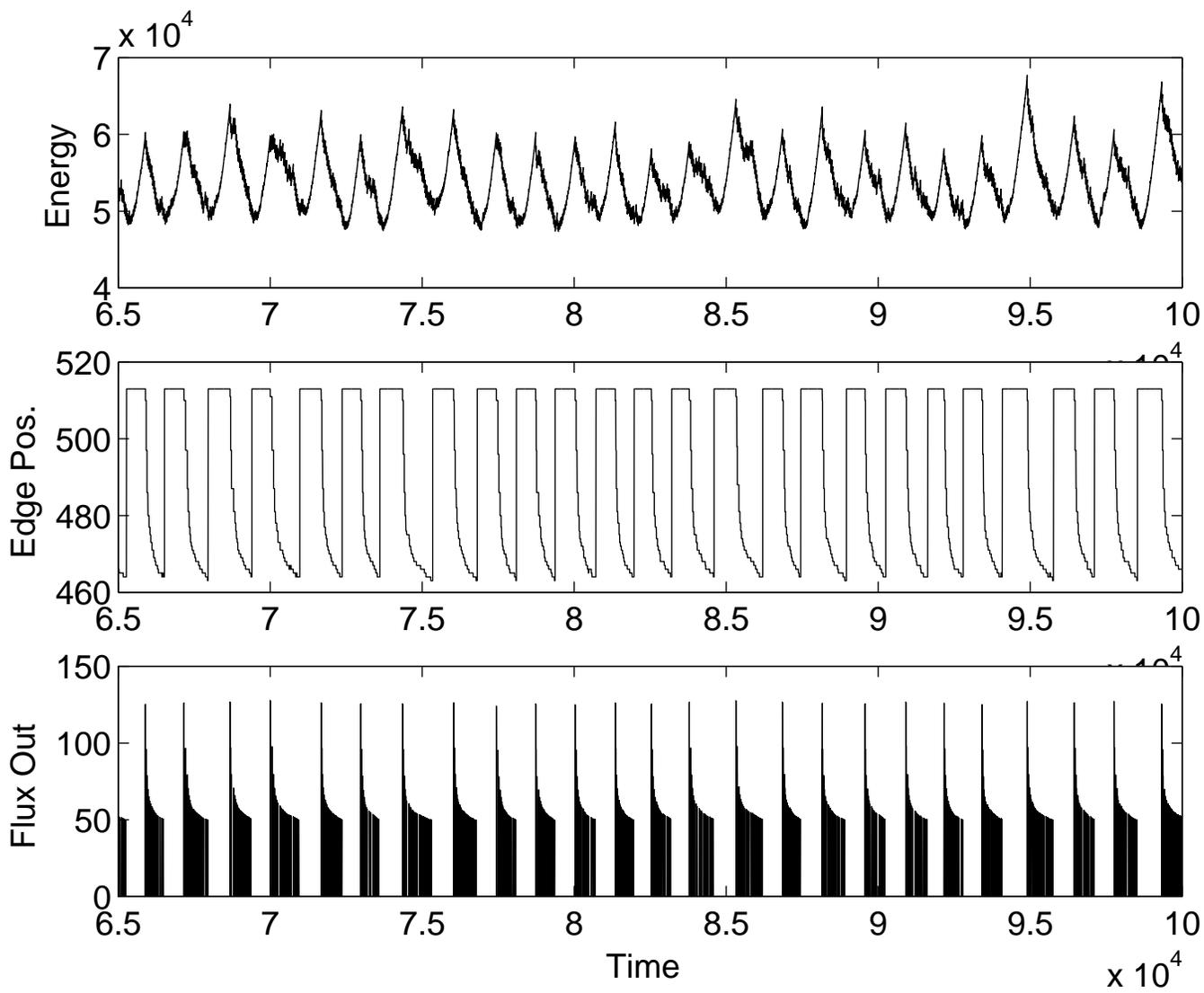
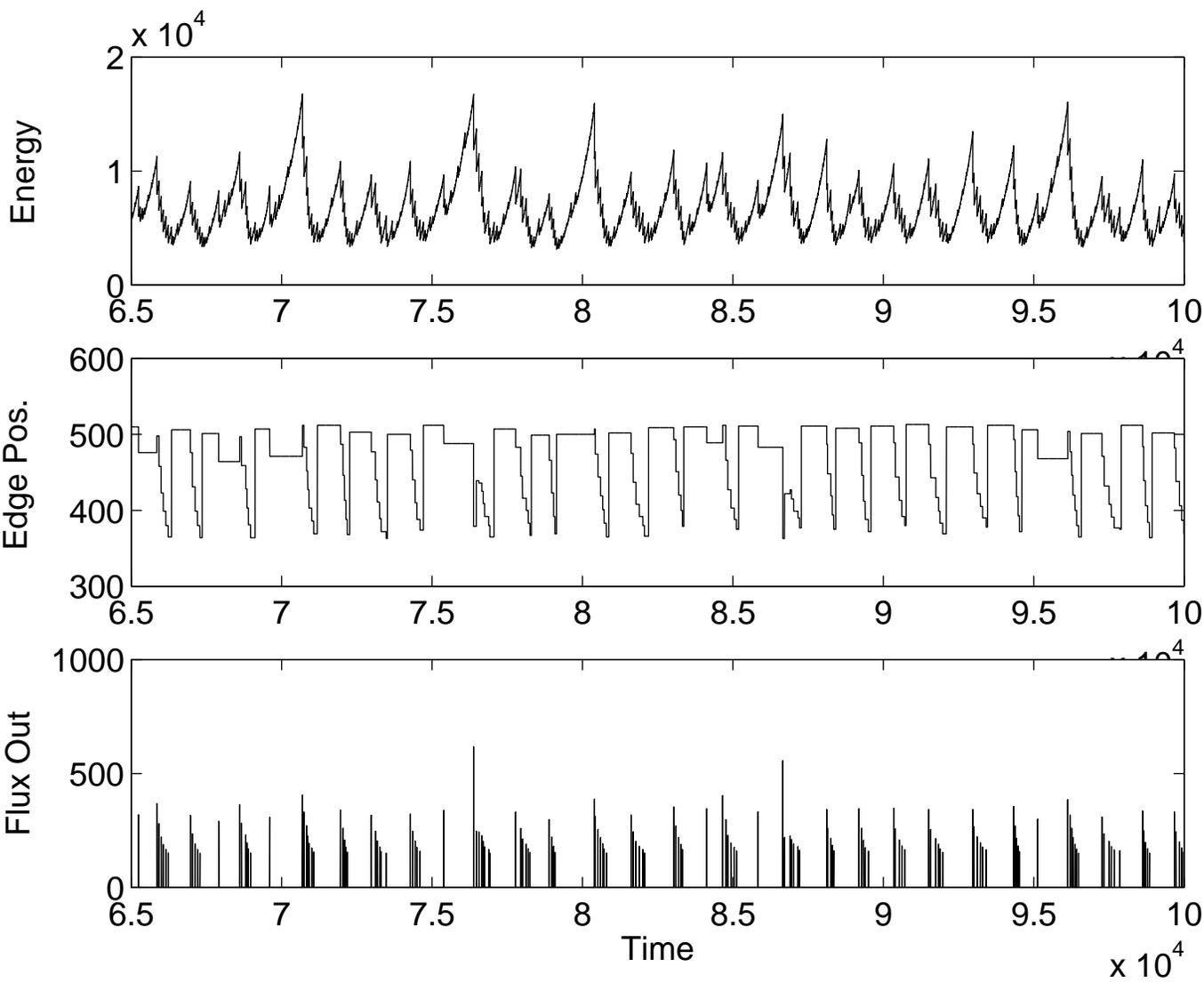
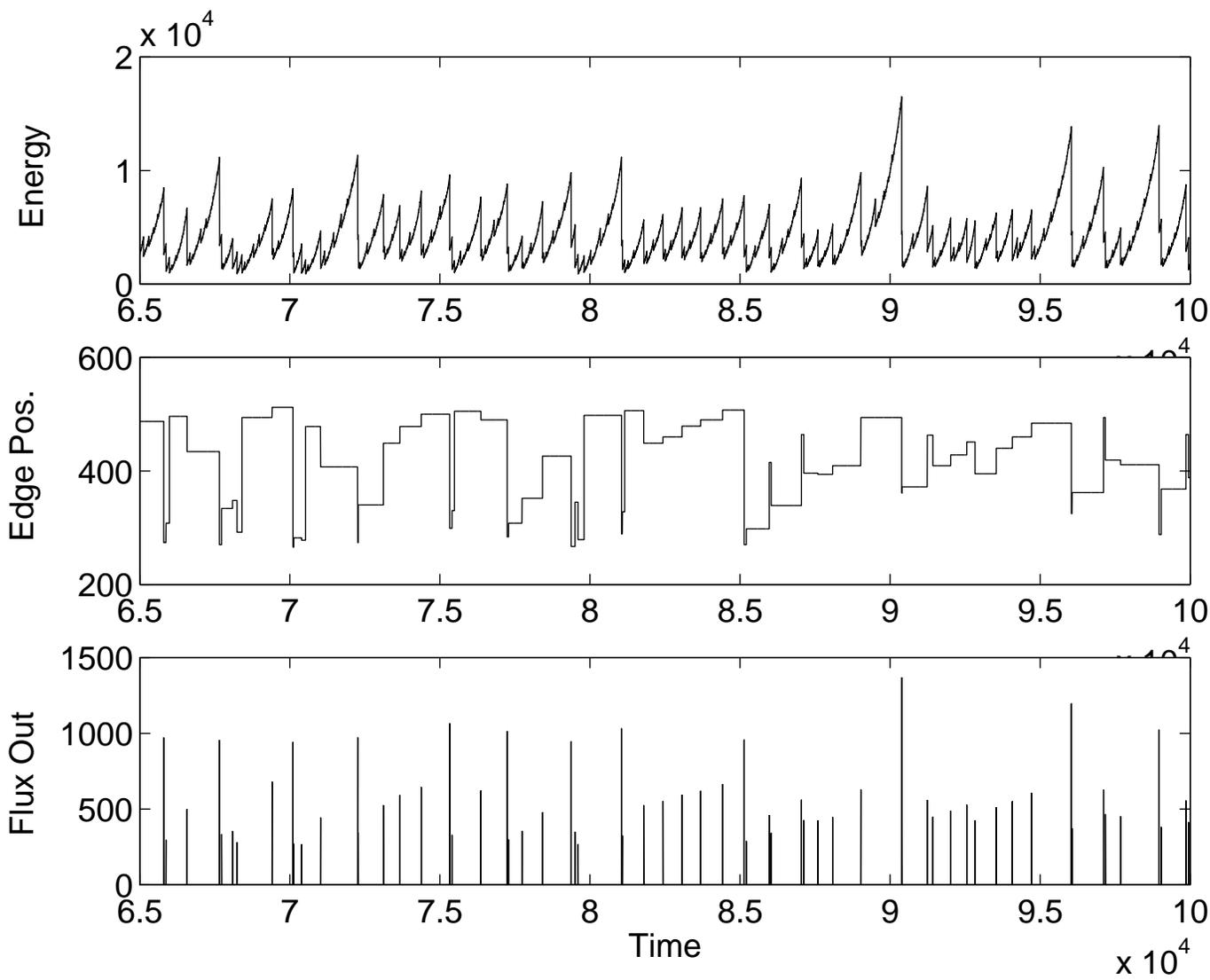
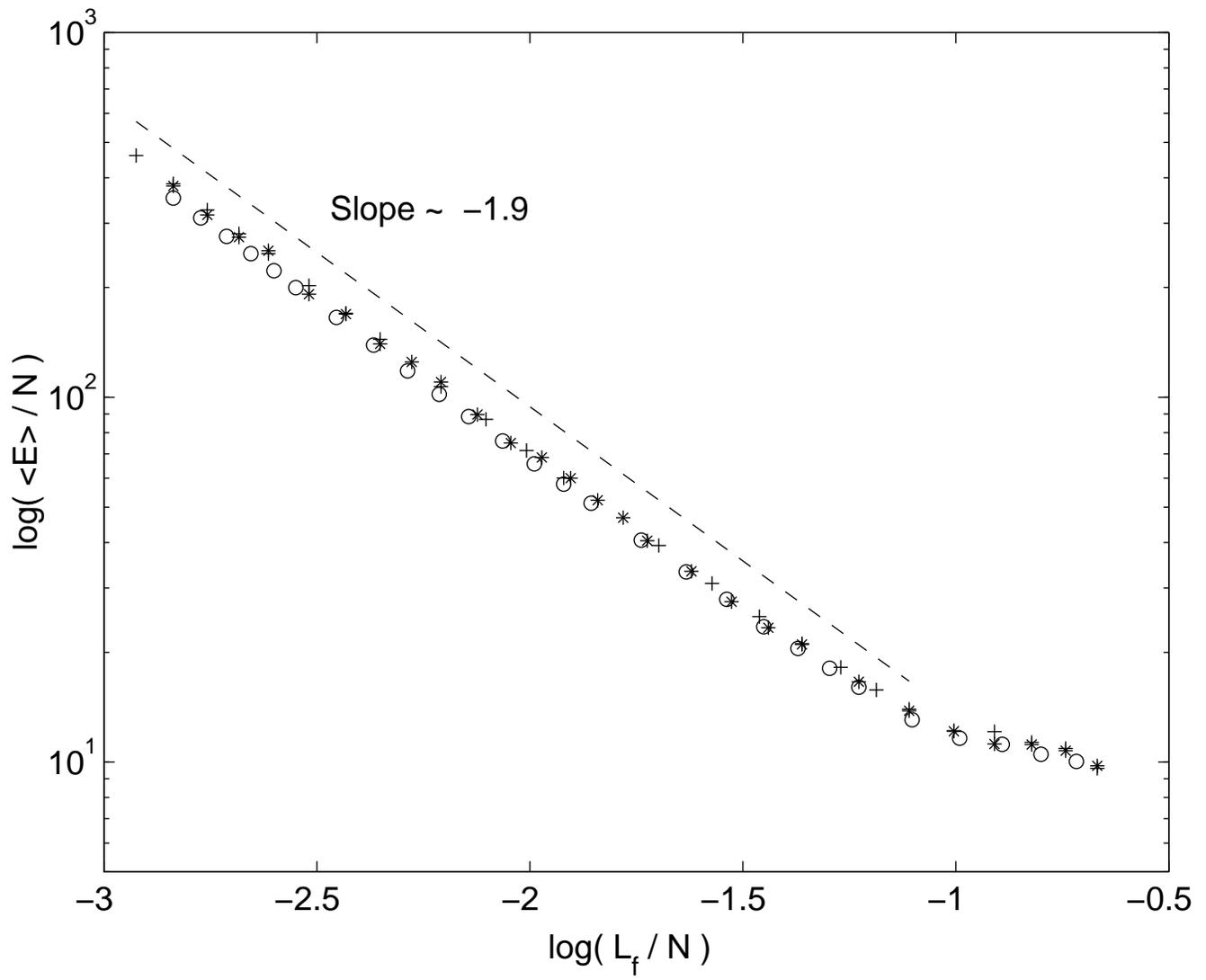
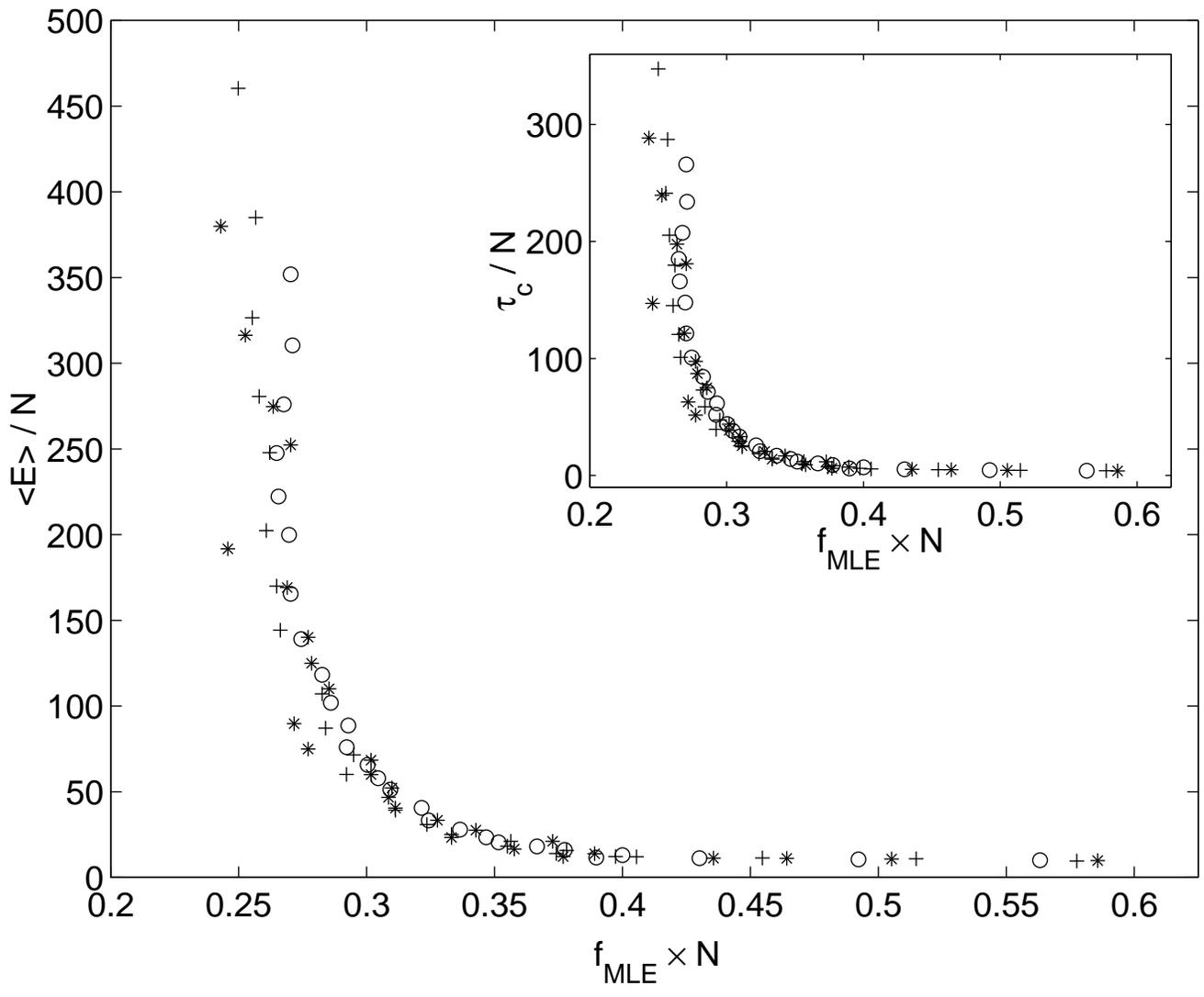
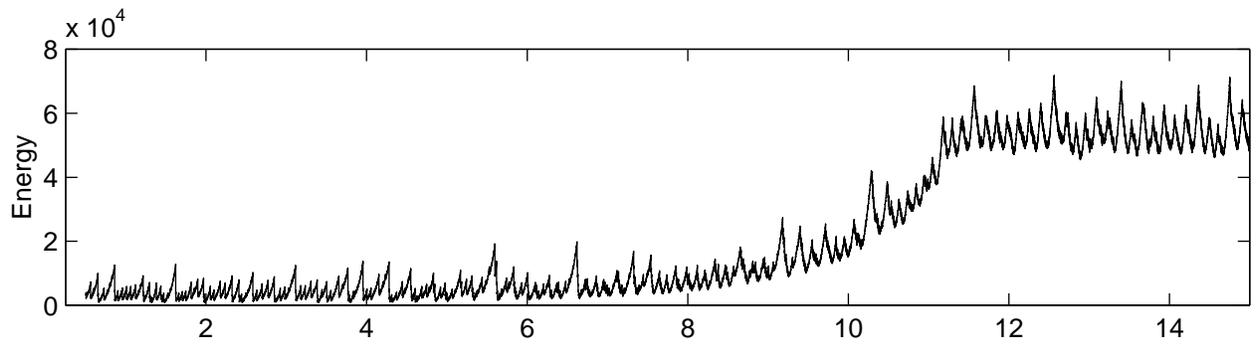
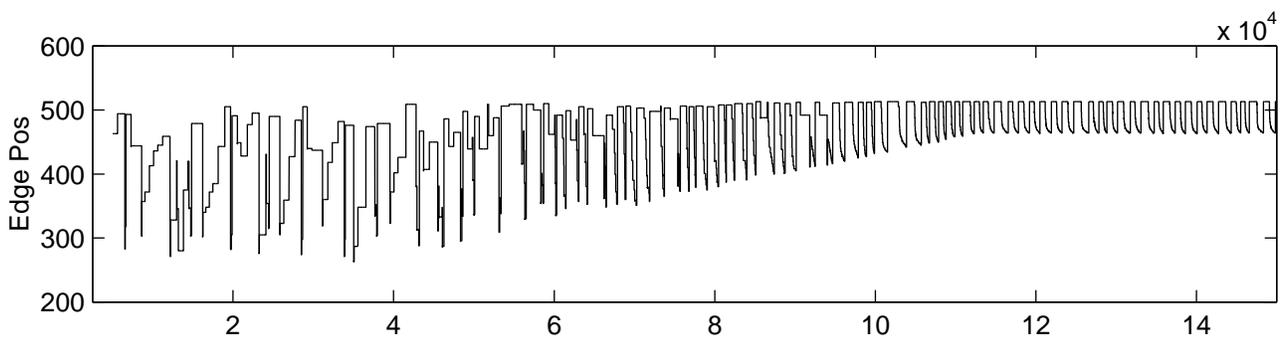
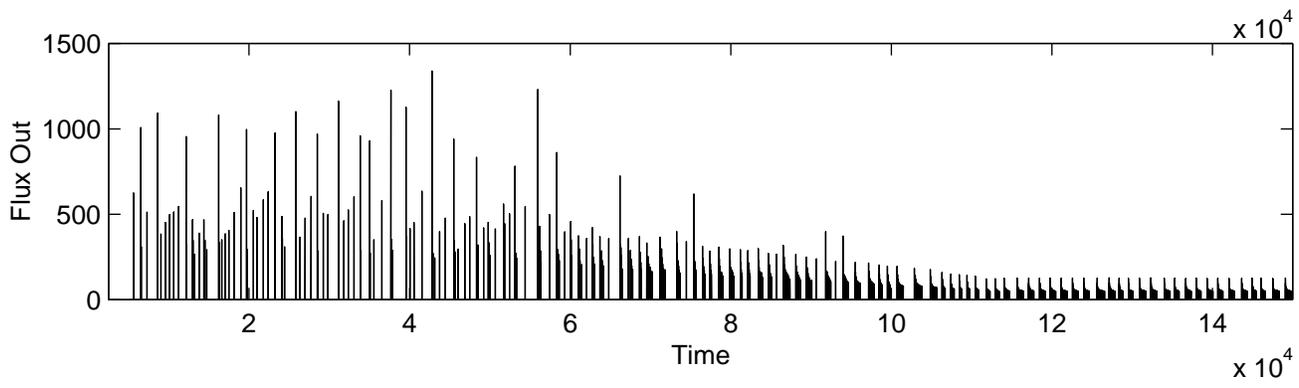
\end{document}